\documentclass[aps,prd,notitlepage,nofootinbib,superscriptaddress]{revtex4-1}
\usepackage[utf8]{inputenc}
\usepackage[table ]{ xcolor}

\usepackage{multirow}

\usepackage{hyperref}
\usepackage{graphicx}
\usepackage{mathtools}
\usepackage{amsmath,amssymb,amsfonts,amsthm,latexsym,stmaryrd,physics,mathrsfs}
\allowdisplaybreaks[4]
\usepackage{marginnote}
\usepackage{color}
\usepackage{bm}
\usepackage{float}
\usepackage{nicefrac}
\usepackage{microtype} 
\usepackage{booktabs}
\usepackage{verbatim}
\usepackage{setspace}
\usepackage[T1]{fontenc}
\usepackage[utf8]{inputenc}
\usepackage{stfloats}
\usepackage{diagbox}
\usepackage{dsfont}
\usepackage{tikz}
\usepackage{geometry}

\hypersetup{
	colorlinks=true
,urlcolor=blue
,anchorcolor=blue
,citecolor=blue
,filecolor=blue
,linkcolor=blue
,menucolor=blue
,pagecolor=blue
,linktocpage=true
,pdfproducer=medialab
}

\def\beq{\begin{equation}}
\def\eeq{\end{equation}}
\newcommand{\bea}{\begin{eqnarray}}
\newcommand{\eea}{\end{eqnarray}}
\def\bi{\begin{itemize}}
\def\ei{\end{itemize}}
\def\ba{\begin{array}}
\def\ea{\end{array}}
\def\bfig{\begin{figure}}
\def\efig{\end{figure}}

\def\be{\begin{eqnarray}}
\def\ee{\end{eqnarray}}

\begin{document}
\raggedbottom
\title{Connection Dynamics of Reduced 5-dimensional Kaluza-Klein Theory and Its Deparametrization}

\author{Haida Li} 
\email{haidali@mail.bnu.edu.cn}
\affiliation{Department of Physics, Beijing Normal University, Beijing 100875, China}

\author{Shengzhi Li}
\email{201921140014@mail.bnu.edu.cn}
\affiliation{Department of Physics, Beijing Normal University, Beijing 100875, China}

\author{Yongge Ma}
\email{mayg@bnu.edu.cn}
\affiliation{Department of Physics, Beijing Normal University, Beijing 100875, China}

\begin{abstract}
    The connection dynamics of the 5-dimensional Kaluza-Klein theory reduced on 4-dimensional
    spacetime is obtained by performing the Hamiltonian analysis and canonical transformations. Deparametrization is achieved in the spherically symmetric model of the theory without introducing additional matter fields beyond the 5-dimensional gravity. Thus the physical time evolution and the physical spatial coordinate can be provided by the geometrical degrees of freedom in the higher dimensional spacetime.
\end{abstract}

\maketitle

\section{Introduction}

In the past three decades, the problem of defining physical time evolution in canonical loop quantum gravity (LQG) has been tackled. The main problem begins with the Hamiltonian formalism of general relativity (GR), where the physical time evolution of gauge invariant Dirac observables is difficult to be realized due to their Poisson commutativity with the Hamiltonian of the system, i.e., $\{O_D,H\}=0$. This implies a trivial physical time evolution for those Dirac observables. Early works on this topic are mainly based on Rovelli's relational framework \cite{rovelli1,rovelli2,rovelli3,rovelli4} as well as the Brown-Kuchař mechanism \cite{brown1995dust} where dust fields are introduced to the Hamiltonian system of GR to serve as reference fields of both space and time. The relational framework was later on improved and applied to LQG in Refs. \cite{dittrich2006partial,dittrich2007partial}, and in a series of works the detailed construction of gauge invariant Dirac observables was achieved for GR couple to dust fields \cite{giesel2010algebraic,giesel2010manifestly,giesel2015scalar}, where some dust fields have been considered as reference fields to achieve classical deparametrization of both Hamitonian and diffeomorphism constraints of GR, providing a viable approach to define physical time evolution and dynamical spatial coordinates.

It is then natually expected that the relational framework, which solves the problem of defining physical time evolution in the Hamiltonian formalism of GR, will also solve the problem in its quantum theory. In fact, there are at least two approaches developed to derive quantum dynamics from deparametrized canonical LQG. The first approach is the so-called reduce phase space quantization approach proposed in Refs. \cite{thiemann2006reduced,giesel2010algebraic}. In this program, Dirac observables related to phase space variables are first constructed at classical level by deparametrizing the first-class constraints using the degrees of freedom provided by some dust fields, leading to the reduction of phase space where only physical degrees of freedom remain for the classical theory. Meanwhile, physical time evolution of the constructed Dirac observables can be given as their relational evolution with respect to the dust fields. Thus the anomaly arising from quantum constraint algebra is no longer an issue since all constraints are imposed at classical level. Also, it has been demonstrated that both the Hamiltonian and diffeomorphism constraints can be deparametrized. This leads to a fully deparametrized system, where not only physical time evolution are defined for Dirac observables, but also their physical spatial coordinates. The second approach towards defining physical time evolution in LQG is introduced in Refs. \cite{domagala2010gravity,lewandowski2016loop}, in which the relational evolution of gravity is defined with respect to a massless scalar field at quantum level. Being different from the first approach, where constraints are imposed at classical level, in the second approach the quantization is first performed to the Hamiltonian constrained system of gravity coupled to the massless scalar field. Solutions of the quantum constraint equation restrict the kinematical Hilbert space to the physical Hilbert space, and in the Heisenberg picture the physical time evolution of operators in the kinematical Hilbert space is obtained by the construction of their corresponding quantum Dirac observables using quantum evolution operator defined on the physical Hilbert space. The models of canonical LQG following this approach have produced some very interesting results, including the bouncing behaviour at full quantum level \cite{zhang2019bouncing}, which reveals the resemblance between full LQG and its symmetry-reduced theories such as loop quantum cosmology (LQC), where the quantum bounce is predicted as the resolution of the big bang singularities in GR \cite{ashtekar2006quantum,ashtekar2006quantum123,ding2009effective,yang2009alternative}. However, one of the challanges to this approach is that at quantum level the diffeomorphism constaint is solved by using the group-averaging technique while the Hamiltonian constraint is deparametrized. To obtain a consistent treatment to both constraints, how to implement physical spatial coordinates in this approach remains an open issue.

Despite the fact that there has been much success in solving the problem of time in LQG, in current results one only considered the relational evolution with respect to certain matter fields. In this paper, our main purpose is to consider the systems of higher dimensional gravity, where the degrees of freedom (DoFs) chosen to fully deparametrize the system are provided by spacetime geometry itself. Conceptually speaking, in such a system physical time evolution is not given by some additionally introduced matter fields, instead it is depicted by spacetime degrees of freedom including the extra dimensions.

Higher dimensional gravity represents a series of efforts to unify 4-dimensional gravity and other interactions. We will consider the 5-dimensional Kaluza-Klein theory, which can be reduced on 4-dimensional spacetime as gravity coupled with an electromagnetic (EM) field and a scalar field \cite{kaluza1921unitatsproblem,klein1926quantum,bailin1987kaluza}. Note that if the extra dimension is compactified as a circle with a microscopic radius, a killing vector field may arise naturally in the low energy regime \cite{blagojevic2002gravitation}. The killing reduction of the 5-dimensional gravity has been studied previously \cite{yang68killing}, where the 5-dimensional spacetime is reduced along the trajectory of the killing vector along the fifth dimension so that a 4-dimensional gravity with some extra DoFs from the reduced fifth dimension is generated. Meanwhile, following the same idea on the extension of loop quantization to some modified gravity theories \cite{zhang2011loop,zhang2011extension,zhang2011nonperturbative}, the killing reduced theory is for the first time recast in connection formalism in this paper such that the study of its loop quantization becomes possible. However, it is rather difficult in the full 4-dimensional theory to deparametrize both Hamiltonian and diffeomorphism constraints simultaneously. As a simplified model, spherically symmetric reduction will be applied to the theory in the connection formalism such that a fully deparametrized theory where both physical time evolution and physical spatial coordinates for gauge invariant complete observables can be well defined.

The structure of this paper is organised as the following. In section II we briefly review the classical framework of defining relational evolution in Hamiltonian constrained systems. In section III we perform Hamiltonian analysis of the 5-dimensional Kaluza-Klein theory reduced on 4-dimensional spacetime, obtain the connection dynamics and then perform spherically symmetric reduction to the theory. In section IV we will perform deparametrization to the spherically symmetric model, and both Hamiltonian constraint and diffeomorphism cosntraint will be deparametrized, providing relational physical time evolution and physical spatial coordinates to the system. In the last section we will summarize the results and provide some outlook towards future prospects in this direction. 

\section{Relational Framework of GR coupled to matter fields}

In the scheme to establish the relational framework for GR couple to dust fields and massless scalar field, one usually begins with identifying the DoFs used for representing relational evolution. Then the constrained system is reformulated into its deparametrization form, enabling the construction of complete observables which are invariant along the gauge orbit induced by the first-class constraints. The exact procedure of deparametrization is closely related to the specific form of constraints.

Consider GR coupled to matter fields in general cases on a manifold $M$ with proper foliation $\Sigma \times \mathcal{R}$. Suppose that one could identify two sets of canonical pairs $(P^I,T_I)$ and $(q^a,p_a)$ representing the reference fields and remaining DoFs respectively, such that the constraints can be written in the equivalent form:
\begin{equation}
    \begin{split}
        C_I(x)=P_I(x)+h_I[T^I,q^a,p_a](x),
    \end{split}
\end{equation}
where $C_I$ denotes all first-class constraints including the Hamiltonian and spatial diffeomorphism constraints. Note that in comparison to the cases of GR coupled to the dust fields and the massless scalar field treated before, we admit the physical coordinates $T^I$ appearing in the expression of $h_I$ in more general cases. Gauge invariant Dirac observables can be defined as follows. Let $F$ be an arbitrary function on the phase space which depends only on $(q^a,p_a)$. Then an observable which is invariant under the gauge transformation of a first-class constraint $C_i$ can be defined as \cite{giesel2010algebraic}:
\begin{equation}\label{observe1}
    O_F(\tau_i):=\left(e^{\{C_i[\beta],.\}}\cdot F\right)_{\beta=\tau^i-T^i},
\end{equation}
where $C_i[\beta]$ represents the smeared version of $C_i$ defined as $C_i[\beta] := \int d^3x\beta(x)C_i(x)$, and the action of $e^{\{C_i[\beta],.\}}$ on $F$ is defined by:
\begin{equation}
    \begin{split}
        e^{\{C_i[\beta],.\}}\cdot F&:=F+\sum\limits^{\infty}_{n=1}\int d^3 x_1...\int d^3 x_n[\beta(x_1)]..[\beta(x_{n})]\{C_i(x_1),\{..\{C_i(x_n),F\}..\}\}.\\
    \end{split}
\end{equation}
Now let us consider a complete gauge invariant observable of both Hamiltonian and diffeomorphism constraints, which can be constructed as \cite{dittrich2006partial}:
\begin{equation}\label{complete}
    \begin{split}
        O_F(\tau,\sigma_j)=O_{O_F(\sigma_j)}(\tau).
    \end{split}
\end{equation}
The following properties can be derived as is the case for GR coupled to dust:
\begin{equation}\label{property}
    \begin{split}
        &\{O_F[\tau,\sigma_j],O_{F'}[\tau,\sigma_j]\}=\{O_F[\tau,\sigma_j],O_{F'}[\tau,\sigma_j]\}^*=O_{\{F,F'\}^*}[\tau,\sigma_j],\\
        &O_{F+F'}[\tau,\sigma_j]=O_F[\tau,\sigma_j]+O_{F'}[\tau,\sigma_j],\qquad O_{F\cdot F'}[\tau,\sigma_j]=O_F[\tau,\sigma_j]\cdot O_{F'}[\tau,\sigma_j],
    \end{split}
\end{equation}
where the Dirac bracket is defined as:
\begin{equation}
    \{f,g\}^*:=\{f,g\}-\sum\limits_{a,b}\{f,\tilde{\phi}_a\}M^{-1}_{ab}\{\tilde{\phi}_b,g\},
\end{equation}
where $\tilde{\phi}_a$, $\tilde{\phi}_b$ are second-class constraints and $M_{ab}=\{\tilde{\phi}_a,\tilde{\phi}_b\}$. The relational evolution of Dirac observable $O_F(\tau,\sigma_j)$, namely the first order derivative of $O_F(\tau,\sigma_j)$ with respect to the physical time parameter $\tau$, can thus be calculated as:
\begin{equation}\label{evolution}
    \begin{split}
        \frac{d}{d\tau}O{}_{F}(\tau,\sigma_j)&=\sum\limits^{\infty}_{n=1}\int d^3 \sigma_1'...\int d^3 \sigma_n'[\tau-T(\sigma_1')]..[\tau-T(\sigma_{n-1}')]\{C(\sigma_1'),\{..\{C(\sigma_n'),O_F(\sigma_j)\}..\}\}\\
        &=O{}_{\{h[1],F\}}(\tau,\sigma_j)\\
        &=\{O{}_{(h[1])}(\tau,\sigma_j),O{}_{F}(\tau,\sigma_j)\},
    \end{split}
\end{equation}
where $T(x)$ is a scalar field and $T(\sigma):=O_T(\sigma_j)$. Here it is worth noting that if the function $F$ depended not only on $(q,p)$ but also on $T(x)$. The second step of (\ref{evolution}) may not hold.

From Eq.(\ref{property}) one can see that the multi-parameter family of maps $O^{(\tau,\sigma)}: F\mapsto O_F(\tau,\sigma)$ are actually homomorphisms from the Poisson algebra of functions on phase space to the Poisson algebra of weak Dirac observables. Thus for a general functional:
\begin{equation}
    F = F[q_{a}(x),p^{a}(x),T^I(x),P_I(x)],
\end{equation}
we have the following identity:
\begin{equation}\label{ftof}
    O_F(\tau,\sigma_j) =  F[O_{q_{a}}(\tau,\sigma_j),O_{p^{a}}(\tau,\sigma_j),\tau,\sigma_j,-O_{h}(\tau,\sigma_j),-O_{h_j}(\tau,\sigma_j)].
\end{equation}
Using this identity, phase space functions can be easily constructed as gauge invariant Dirac observables on the reduced phase space and the relational framework is thus established.

\section{Hamiltonian Formulation of 5D Kaluza-Klein theory}

As is the purpose of this work, we will consider 5-dimensional Kaluza-Klein gravity which in itself can provide degrees of freedom for deparametrization, since the refernece fields can be provided without introducing additional fields. In this section, we will first perform the Hamiltonian analysis to the Kaluza-Klein theory reduced on 4-dimensional spacetime. Then, we will transform the Hamiltonian theory canonically into connection formalism suitable for loop quantization. Finally, to carry out the deparametrization, the spherically symmetric model of the theory will also be studied. 

\subsection{Hamiltonian Analysis of 5D Kaluza-Klein Theory}

Since the extra dimension is compactified as a circle with a microscopic radius in 5-dimensional Kaluza-Klein theory, a killing vector field naturally arise in the low energy regime. Hence one may parametrize the 5-dimensional metric by the 4-dimensional fields as \cite{bailin1987kaluza}:
    \begin{equation}
        \begin{split}
            &g_{55}=\phi^2\\
            &g_{5\mu}=g_{\mu 5}=\phi^2\mathcal{A}_{\mu}\\
            &g_{\mu\nu}=h_{\mu\nu}+\phi^2\mathcal{A}_{\mu}\mathcal{A}_{\nu},
        \end{split}
    \end{equation}
where $\mu$, $\nu=0,1,2,3$, $h_{\mu\nu}$ represents a 4-dimentioanl metric, $\mathcal{A}_{\mu}$ can be identified as an EM 4-potential, and $\phi$ is a scalar field. The action for 5-dimensional Kaluza-Klein theory reads:
\begin{equation}
    S=\frac{1}{\kappa_5}\int_{\mathcal{M}^5} d^5x\sqrt{|g|}({}^5R),
\end{equation}
where $^5R$ denotes the Ricci scalar in five dimensions. This action can be reduced to four dimensions up to a boundary term as:
\begin{equation}\label{action1}
    S = \frac{1}{\kappa}\int_{\mathcal{M}^4}d^4x\phi\sqrt{-h}(R+\frac{1}{4}\phi^2\mathcal{F}_{\mu\nu}\mathcal{F}^{\mu\nu}),
\end{equation}
where $\kappa$ is a constant defined as $\frac{1}{\kappa}:=\frac{1}{\kappa_5}\int dx^5$ such that the integration in the fifth dimension is carried out, $\phi$ is set to be positive, $R$ is the 4-dimentional scalar curvature determined by $h_{\mu\nu}$, and $\mathcal{F}_{\mu\nu}=(d\mathcal{A})_{\mu\nu}$ is the EM tensor.

By the standard 3+1 decomposition of the reduced 4-dimensional spacetime, the components of the 4-metric can be written as:    
\begin{equation}
    \left(h_{\mu\nu}\right)=\left(                 
    \begin{array}{cc}   
      -N^2+q_{ab}N^aN^b & N_a\\  
      N_a & q_{ab}\\  
    \end{array}
  \right),\quad a,b=1,2,3   
\end{equation}
where $N_a=h_{ab}N^b$. It's inverse reads:
\begin{equation}
    \left(h^{\mu\nu}\right)=\left(                 
    \begin{array}{cc}   
      -\frac{1}{N^2} & \frac{N^a}{N^2}\\  
      \frac{N^b}{N^2} & q^{ab}-\frac{N^aN^b}{N^2}\\  
    \end{array}
  \right).
\end{equation}
Following the treatment in \cite{lacquaniti2006adm}, the 4-dimensional scalar curvature $R$, the extrinsic curvature $K_{ab}$ of the 3-dimensional spatial hypersurface and the EM tensor take the following form:
    \begin{equation}
        \begin{split}
            R&=(^3R+K_{ab}K^{ab}-K^2)+2\nabla_a(-n^b\nabla_bn^a+n^aK),\\
            K_{ab}&=\frac{1}{2N}(\dot{q}_{ab}-D_aN_b-D_bN_a),\\
            \mathcal{F}_{\mu\nu}\mathcal{F}^{\mu\nu}&=\mathcal{F}_{ab}\mathcal{F}^{ab}-\frac{2}{N^2}q^{ab}M_aM_b,\\
        \end{split}
    \end{equation}
where $M_a:=\mathcal{F}_{a0}-N^b\mathcal{F}_{ab}$, $\nabla_{\mu}$ is the covariant derivative for the 4-dimensional spacetime, $n^{\mu}$ and $D_{a}$ are respectively the normal vector and the covariant derivative of the 3-dimensinal hypersurface. Furthermore, it is worth noting that the total derivative terms contained in scalar curvature can not be neglected due to the non-minimally coupled scalar field $\phi$ in the action (\ref{action1}). This term can be written as \cite{lacquaniti2006adm}:
\begin{equation}
    \begin{split}
        &\sqrt{-h}\phi\nabla_{a}(n^{a}K-n^{b}\nabla_bn^{a})=-N\sqrt{q}K\partial_n\phi-\sqrt{q}ND^{a}D_{a}\phi,
    \end{split}
\end{equation}
where we used the notation $\partial_n\phi:=n^a\nabla_a\phi=(t^a-N^a)/N\nabla_a\phi=\frac{1}{N}(\partial_0\phi-N^a\partial_a\phi)$. The Langrangian density in (\ref{action1}) can thus be rewritten in terms of the spatial variables up to boundary terms as:
    \begin{equation}
        \begin{split}
            \mathcal{L}=\frac{N}{\kappa}\sqrt{q}\phi(^3R+K_{ab}K^{ab}-K^2)-\frac{2}{\kappa}N\sqrt{q}(q^{ab}D_a\partial_b\phi+K\partial_n\phi)+\frac{1}{4\kappa}N\sqrt{q}\phi^3(\mathcal{F}_{ab}\mathcal{F}^{ab}-\frac{2}{N^2}q^{ab}M_aM_b).
        \end{split}
    \end{equation}
By the Legendre transformation, there are six pairs of conjugate variables $(N,\pi_N)$, $(N^a,\pi_{N^a})$, $(\mathcal{A}_0,\pi_{\mathcal{A}_0})$, $(q_{ab},\Sigma^{ab})$, $(\mathcal{A}_a,\pi^a)$ and $(\phi,\pi_{\phi})$, satisfying:
\begin{equation}\label{momentum}
    \begin{split}
        &\pi_N=\frac{\partial\mathcal{L}}{\partial\dot{N}}=0,\pi_{N^a}=\frac{\partial\mathcal{L}}{\partial\dot{N}^a}=0,\pi_{A_0}=\frac{\partial\mathcal{L}}{\partial\dot{A}_0}=0,\\
        &\Sigma^{ab}=\frac{\partial\mathcal{L}}{\partial\dot{q}_{ab}}=\frac{1}{\kappa}(\phi \sqrt{q}(K^{ab}-Kq^{ab})-\sqrt{q}q^{ab}\partial_n\phi),\\
        &\pi^a=\frac{\partial\mathcal{L}}{\partial\dot{A}_{a}}=\frac{1}{\kappa N}\sqrt{q}\phi^3q^{ab}M_b,\\
        &\pi_{\phi}=\frac{\partial\mathcal{L}}{\partial\dot{\phi}}=-\frac{2}{\kappa}\sqrt{q}K.
    \end{split}
\end{equation}
It can be seen from (\ref{momentum}) that $\pi_N=0$, $\pi_{N^a}=0$ and $\pi_{\mathcal{A}_0}=0$ are three sets of primary constraints, and their corresponding configuration variables are merely Lagrangian multipliers. The non-vanishing Poisson brackets of phase space variables read:
\begin{equation}
    \begin{split}
        &\{q_{ab}(x),\Sigma^{cd}(y)\}=\delta^{(c}_a\delta^{d)}_b\delta^{(3)}(x,y),\\
        &\{\phi(x),\pi_{\phi}(y)\}=\delta^{(3)}(x,y),\\
        &\{\mathcal{A}_a,\pi^a\}=\delta^{(3)}(x,y).
    \end{split}
\end{equation}
The Hamiltonian density can be derived as $\mathcal{H}=\dot{q}^ip_i-\mathcal{L}=NH+N^aV_a+\mathcal{A}_0J$, with the constraints \cite{lacquaniti2006adm}:
\begin{equation}\label{constraints1}
    \begin{split}
        H=&-\frac{\phi}{\kappa}\sqrt{q}R+\frac{2}{\kappa}\sqrt{q}q^{ab}D_a\partial_b\phi+\frac{\kappa N}{\sqrt{q}\phi}(\Sigma_{ab}\Sigma^{ab}-\frac{\Sigma^2}{2}+\frac{(\Sigma-\phi\pi^{\phi})^2}{6})\\
        &-\frac{1}{4\kappa}\phi^3\sqrt{q}\mathcal{F}_{ab}\mathcal{F}^{ab}-\frac{\kappa}{2\phi^3\sqrt{q}}\pi^a\pi^bq_{ab},\\
        V_a=&-2q_{ab}D_c\Sigma^{cb}+\pi_{\phi}\partial_a\phi-\pi^b\mathcal{F}_{ab}+\mathcal{A}_a\partial_b\pi^b,\\
        J=&-\partial_a\pi^a,
    \end{split}
\end{equation}
where the term $\mathcal{A}_a\partial_b\pi^b$ proportional to the constraint $J$ has been added into $V_a$ so that it generates the correct spatial diffeomorphism trnasformations for $(\mathcal{A}_a,\pi^a)$. Thus $H$ is the Hamitonian constraint associated with time evolution, $V_a$ is the spatial diffeomorphism constraint which generates spatial diffeomorphism transformations, $J$ is the Gaussian constraint for the EM field. It can be further verified that all the secondary constraints in (\ref{constraints1}) are of first class. Especially, for the smeared secondary constraints one has:
\begin{equation}
    \begin{split}
        &\{J[\mathcal{A}_0],J[\mathcal{B}_0]\}=0,\quad \{J[\mathcal{A}_0],V[\vec{N}]\}=-J(\mathscr{L}_{\vec{N}}\mathcal{A}_0),\quad \{J[\mathcal{A}_0],H[N]\}=0,\\
        &\{V[\vec{N}],V[\vec{N}']\}=V([\vec{N},\vec{N}']),\quad \{H[M],V[\vec{N}]\}=-H(\mathscr{L}_{\vec{N}}M),\\
        &\{H[N],H[M]\}=V(ND^aM-MD^aN),
    \end{split}
\end{equation}
where $J[\mathcal{A}_0]=\int_{\Sigma}d^3x \mathcal{A}_0(x)J(x)$, $H(N)=\int_{\Sigma} d^3xN(x)H(x)$ and $V(\vec{N})=\int_{\Sigma} d^3xN^a(x)V_a(x)$.

\subsection{Connecton Dynamics}

Recall that the non-perturbative loop quantization of GR is based on its connection dynamical formalism. In this subsection we will derive a connection dynamical formalism for the 5D Kaluza Klein theory in 4-dimensional spacetime. In order to achieve this goal, we first define:
\begin{equation}
    \tilde{K}^{ab}:= \phi K^{ab}+\frac{q^{ab}}{2N}(\dot{\phi}-N^c\partial_c\phi).
\end{equation}
This leads to the simplification of the conjugate momentum of $q_{ab}$ as:
\begin{equation}
    \begin{split}
        \Sigma^{ab}=\frac{1}{\kappa}(\tilde{K}^a_iE^{bi}-\frac{1}{q}\tilde{K}^i_cE^c_iE^a_jE^b_j),
    \end{split}
\end{equation}
where $\tilde{K}^a_i:= \tilde{K}^{ab}e^i_b$, and $E^a_i=\sqrt{q}e^a_i$ is the densitized triad. Then the following property is guaranteed:
\begin{equation}
     \begin{split}
        \int d^4 \Sigma^{ab}\dot{q}_{ab}=\frac{2}{\kappa}\int d^4 E^a_i\frac{d\tilde{K}^i_a}{dt},
     \end{split}
\end{equation}
and the Poisson structure of $(\tilde{K}^i_a,E^b_j)$ can be calculated using the original Poisson structure as:
\begin{equation}
    \begin{split}
        \{K^i_a(x),K^j_b(y)\}&=0,\\
        \{E^a_i(x),E^b_j(y)\}&=0,\\
        \{K^i_a(x),E^b_j(y)\}&=\frac{\kappa}{2}\delta^b_a\delta^i_j\delta(x,y).
    \end{split}
\end{equation}
We can canonically transform the Hamiltonian system using the new sets of variables $(\tilde{K}^i_a,E^b_j)$, $(A_a,\pi^b)$, $(\phi,\pi_{\phi})$. Since $\tilde{K}^{ab}$ is symmetric, i.e. $\tilde{K}^{ab}=\tilde{K}^{ba}$, an additional set of constraints are introduced:
\begin{equation}
    \begin{split}
        &G_{jk}:=\tilde{K}_{a[j}E^a_{k]}=0,\qquad j,k=1,2,3\\
    \end{split}
\end{equation}
which is equivalent to the following constraint:
\begin{equation}\label{gauss}
    G_j(x)=\epsilon_{jkl}\tilde{K}^k_aE^a_l.
\end{equation}
The smeared version of this constraint can be written as:
\begin{equation}
    G[\Lambda]=\int_{\Sigma} d^3xG_j(x)\Lambda^j(x).
\end{equation}
Now we define:
\begin{equation}
    \begin{split}
        &A^i_a:=\Gamma^i_a+\gamma\tilde{K}^i_a,\\
    \end{split}
\end{equation}
where $\Gamma^i_a$ is the spin-connection determined by $E^a_i$ and $\gamma$ is a non-zero real number known as the Barbero-Imirzi parameter in LQG \cite{han2007fundamental}. A second canonical transformation can be performed by changing the variables depicting the gravitational degrees of freedom from $(\tilde{K}^i_a,E^b_j)$ to $(A^i_a,E^b_j)$. The Poisson brackets between the new sets of variables can be calculated as:
\begin{equation}
    \begin{split}
        &\{E^{a}_i(x),E^{b}_j(y)\}=0,\\
        &\{A^i_{a}(x),A^{j}_b(y)\}=0,\\
        &\{A^i_{a}(x),E^{b}_j(y)\}=\frac{\kappa\gamma}{2}\delta^{(b}_a\delta^{i)}_j\delta^{(3)}(x,y).\\ 
    \end{split}
\end{equation}
The compatibility condition between the covariant derivative and the densitized triad implies:
\begin{equation}
    \partial_aE^a_j+\epsilon_{jkl}\Gamma^k_aE^a_l=0.
\end{equation}
Then taking account of (\ref{gauss}), the standard Gaussian constraint can be obtained as:
\begin{equation}
    \begin{split}
        G_j&=\mathcal{D}_aE^a_j/\gamma\equiv (\partial_aE^a_j+\epsilon_{jkl}A^k_aE^{al})/\gamma=0,
    \end{split}
\end{equation}
which suggests that $A^i_a$ is a $su(2)$-connection. Thus our canonical transformations share the same structure as that for the connection formalism of GR despite the fact that we used $\tilde{K}^i_a$ instead of the extrinsic curvature during the canonical transformation from $(q_{ab},\Sigma^{ab})$ to $(\tilde{K}^i_a,E^a_i)$.

We can then rewrite all of the constraints of our system by using the connection variables as:
\begin{equation}\label{Connection Hamil}
    \begin{split}
        H&=\phi[F^j_{ab}-(\gamma^2+\frac{1}{\phi^2})\epsilon_{jmn}\tilde{K}^m_a\tilde{K^n_b}]\frac{\epsilon_{jkl}E^a_kE^b_l}{\sqrt{q}\kappa}\\
        &+\frac{2}{3\kappa\phi\sqrt{q}}(\tilde{K}^i_aE^a_i)^2+\frac{2\sqrt{q}}{\kappa}D^aD_a\phi+\frac{\kappa}{6\sqrt{q}}(\pi_{\phi})^2\phi+\frac{2}{3\sqrt{q}}\pi_{\phi}(\tilde{K}^i_aE^a_i)\\
        &-\frac{1}{4\kappa\sqrt{q}^3}\phi^3\mathcal{F}_{ab}\mathcal{F}_{cd}E^a_iE^c_iE^b_jE^d_j-\frac{2\kappa\sqrt{q}}{\phi^3}\pi^a\pi^bE^i_aE^i_b,\\
        V_a&=\frac{2}{\kappa\gamma}F^j_{ab}E^b_j+\pi_{\phi}\partial_a\phi-\pi^b\mathcal{F}_{ab}+\mathcal{A}_a\partial_b\pi^b,\\
        G_j&=\frac{2}{\kappa\gamma}\mathcal{D}_aE^a_j\equiv \frac{2}{\kappa\gamma}(\partial_aE^a_j+\epsilon_{jkl}A^k_aE^{al}),\\
        J&=-\partial_a\pi^a,
    \end{split}
\end{equation}
where $F^j_{ab}:=2\partial_{[a}A^j_{b]}+\epsilon_{jkl}A^k_aA^l_b$ is the curvature of the connection $A_b^j$. As is shown in (\ref{Connection Hamil}), the expressions of the constraints are complicated. For simplicity we will perform spherically symmetric reduction to the 5-dimensional Kaluza-Klein theory on 4-dimensional spacetime and then fully deparametrize the symmetry-reduced model.
\subsection{Spherically Symmetric Reduction}
Previous works on obtaining the spherically symmetric form of the connection and desitized triad were described in several articles, e.g.\cite{bojowald2000symmetry,kastrup1994spherically}. Later on, there has been a few works discussing both classical simplification and loop quantization of the spherically symmetric system\cite{campiglia2007loop,gambini2013loop,gambini2020spherically,chiou2012loop,zhang2020loop,zhang2022loop}. We start by recalling the spherically symmetry-reduced version of the connection \cite{campiglia2007loop,chiou2012loop}:
\begin{equation}
    \begin{split}
        A=A^i_a\tau_idx^a=&A_x(x)\tau_3dx+(A_1(x)\tau_1+A_2(x)\tau_2)d\theta\\
        &+[(A_1(x)\tau_2-A_2(x)\tau_1)\sin\theta+\tau_3\cos\theta]d\varphi,
    \end{split}
\end{equation}
where $x$ represents the radial coordinate of the spherical symmetric space, $\theta$ and $\varphi$ are the angular coordinates, $A_x$, $A_1$ and $A_2$ are functions of $x$ and $\tau_i=-i\sigma_i/2$ with the Pauli matrices $\sigma_i$ being orthonormal generators of $SU(2)$. Correspondingly, the densitized triad takes the form:
\begin{equation}
    \begin{split}
        E=E^a_i\tau^i\partial_a=&E^x(x)\tau_3\sin\theta\frac{\partial}{\partial x}+(E^1(x)\tau_1+E^2(x)\tau_2)\sin\theta\frac{\partial}{\partial\theta}\\
        &+(E^1(x)\tau_2-E^2(x)\tau_1)\frac{\partial}{\partial\varphi},
    \end{split}
\end{equation}
where $E^x$, $E^1$ and $E^2$ are functions of $x$. In the symmetry-reduced theory, under the coordinate transformation $x\rightarrow\bar{x}(x)$, $A_x$ transforms as a scalar density of weight 1, i.e., $\bar{A}_x(\bar{x})=(\partial x/\partial \bar{x})A_x(x)$, while $A_1$ and $A_2$ are scalars. $E^x$ tranforms as a scalar, while $E^1$ and $E^2$ are scalar densties of weight 1. Meanwhile, The conjugate pairs of the EM field $(\mathcal{A}_a,\pi^a)$ and the scalar field $(\phi,\pi_{\phi})$ take the following form in the spherically symmetric model:
\begin{equation}
    \begin{split}
        &\mathcal{A}=\mathcal{A}_adx^a=\mathcal{A}(x)dx,\quad\pi=\pi^a(\frac{\partial}{\partial x^a})=\pi^x(x)\sin\theta\frac{\partial}{\partial x},\\
        &\phi=\phi(x),\quad \pi_{\phi}=P(x)\sin\theta.\\
    \end{split}
\end{equation}
Also, it is worth noting that the EM field tensor $F_{\mu\nu}$ only has $F_{01}=-F_{10}$ as non-zero components so that $F_{ab}=0$. 

In terms of the symmetry-reduced variables, the symplectic structure is given by:
\begin{equation}
    \begin{split}
        \Omega&=\int_{\Sigma}d^3x\frac{2}{\kappa\gamma}\delta E^a_i \wedge \delta A^i_a+\delta\phi\wedge\delta\pi_{\phi}+\delta A_a\wedge\delta \pi^a\\
        &=\int_Idx(\frac{8\pi}{\kappa\gamma}\delta E^x\wedge\delta A_x+\frac{16\pi}{\kappa\gamma}(\delta E^1\wedge \delta A_1+\delta E^2\wedge\delta A_2)+4\pi\delta\phi\wedge\delta P+4\pi\delta \mathcal{A}\wedge\delta \pi^x),
    \end{split}
\end{equation}
which implies:
\begin{equation}
    \begin{split}
        &\{A_x(x),E^x(x')\}=\frac{\kappa\gamma}{8\pi}\delta(x,x'),\quad \{A_1(x),E^1(x')\}=\frac{\kappa\gamma}{16\pi}\delta(x,x'),\\
        &\{A_2(x),E^2(x')\}=\frac{\kappa\gamma}{16\pi}\delta(x,x'),\quad \{\phi(x),P(x)\}=\frac{1}{4\pi}\delta(x,x'),\\
        &\{\mathcal{A}(x),\pi^x(x')\}=\frac{1}{4\pi}\delta(x,x'),\\
    \end{split}
\end{equation}
and all the other Poisson brackets among the basic variables vanish. The field strength 2-form can be written as:
\begin{equation}
    \begin{split}
        F&=(A^2_1+A^2_2-1)\tau_3\sin\theta d\theta\wedge d\varphi-[(A_1'\tau_2-A_2'\tau_1)-A_x(A_1\tau_1+A_2\tau_2)]\sin\theta d\varphi\wedge dx\\
        &+[(A_1'\tau_1+A_2'\tau_2)+A_x(A_1\tau_2-A_2\tau_1)]dx\wedge d\theta,
    \end{split}
\end{equation}
where $A'_1$ denotes the first-order derivative of $A_1$ along the radial direction. In the spherically symmetric case, the Gaussian constraint for the 4-dimensional gravity reduces to:
\begin{equation}\label{c1}
    G[\lambda]=\frac{8\pi}{\kappa\gamma}\int_I dx\lambda(E^x{}'+2A_1E^2-2A_2E^1)=:\int_I dx\lambda(x)\overline{G}_3(x),
\end{equation}
where $\lambda(x):=\Lambda^{3}(x)$. It is not difficult to verify that the constraint (\ref{c1}) no longer generates full SU(2) gauge transformations, instead it only generates U(1) transformation. Based on this fact, we will perform a change of variables to further simplify the theory. First we define $A_{\rho}$ and $E^{\rho}$ by:
\begin{equation}
    \begin{split}
        &A_1=A_{\rho}\cos\beta,\qquad A_2=A_{\rho}\sin\beta,\\
        &E^1=E^{\rho}\cos(\alpha+\beta),\qquad E^2=E^{\rho}\sin(\alpha+\beta).
    \end{split}
\end{equation}
Note that $A_{\rho}$, $E^{\rho}$ and the inner product $(A_1,A_2)\cdot (E^1,E^2)=A_{\rho}E^{\rho}\cos\alpha$ are invariant under the $U(1)$ rotations generated by the constraint (\ref{c1}). Note also that the angle $\alpha(x)$ is gauge independent and $\beta(x)$ is a pure gauge angle. Then we can canonically transform the set of basic variables $(A_x,A_1,A_2,A^e,\phi;E^x,E^1,E^2,\pi^x,\pi^{\phi})$ to the new set of $(A_x,\bar{A}_{\rho},\eta,\mathcal{A},\phi; E^x,E^{\rho},P^{\eta},\pi^x,P)$ by defining:
\begin{equation}
    \begin{split}
        \bar{A}_{\rho}&:=2A_{\rho}\cos\alpha,\\
        \eta&:=\alpha+\beta,\\
        P^{\eta}&:=2A_{\rho}E^{\rho}\sin\alpha=2A_1E^2-2A_2E^1.
    \end{split}
\end{equation}
With those new variables, the Gaussian constraint (\ref{c1}) can be written as:
\begin{equation}
    \overline{G}_3=\frac{8\pi}{\kappa\gamma}(E^x{}'+P^{\eta}).
\end{equation}
Moreover, another change of variables can be achieved to further simplify the theory. Following the same approach introduced in \cite{campiglia2007loop,chiou2012loop}, let:
\begin{equation}
    \begin{split}
        &\bar{A}_x:=A_x+\eta'= A_x+(\alpha+\beta)',\\
        &\bar{P}^{\eta}:=P^{\eta}+E^x{}'.
    \end{split}
\end{equation}
Under this transformation, the Gaussian constraint becomes:
\begin{equation}
    \overline{G}_3=\frac{1}{2G\gamma}(\bar{P}^{\eta}),
\end{equation}
which can be easily solved by setting $\bar{P}^{\eta}=0$ and gauge fixing $\eta=0$. In terms of the new canonical variables $(\bar{A}_x,\bar{A}_{\rho},\eta;E^x,E^{\rho},P_{\eta})$, we can compute the Poisson brackets between phase space variables as:
\begin{equation}
    \begin{split}
        &\{\bar{A}_x(x),E^x(x')\}=\frac{\kappa\gamma}{8\pi}\gamma\delta (x,x'),\quad \{\bar{A}_{\rho}(x),E^{\rho}(x')\}=\frac{\kappa\gamma}{8\pi}\gamma\delta(x,x'),\\
        &\{\phi(x),P(x')\}=\frac{1}{4\pi}\delta(x,x'),\quad \{\mathcal{A}(x),\pi^x(x')\}=\frac{1}{4\pi}\delta(x,x').\\
    \end{split}
\end{equation}
Then our constrained system is simplified to the following three constraints:
\begin{equation}\label{spherical1}
    \begin{split}
        H&=8\pi[\frac{\kappa}{12E^{\rho}\sqrt{|E^x|}}P^2\phi +\frac{1}{3E^{\rho}\sqrt{|E^x|}}P\tilde{K}-\frac{\phi E^{\rho}}{\kappa\sqrt{|E^x|}}-\frac{\bar{A}^2_{\rho}E^{\rho}}{4\kappa\gamma^2\phi\sqrt{|E^x|}}\\
        &\qquad -sgn(E^x)\frac{\sqrt{|E^x|}\bar{A}_{\rho}\bar{A}_x}{\kappa\gamma^2\phi}+sgn(E^x)\frac{\phi\sqrt{|E^x|}}{\kappa}\partial_x(\frac{E^x{}'}{E^{\rho}})+\frac{\phi(E^x{}')^2}{4\kappa\sqrt{|E^x|}E^{\rho}}\\
        &\qquad +\frac{\tilde{K}^2}{3\kappa\phi E^{\rho}\sqrt{|E^x|}}-\frac{\kappa\pi^x\pi^x E^{\rho}}{\phi^3\sqrt{|E^x|}|E^x|}+\frac{E^{\rho}\sqrt{|E^x|}}{\kappa}D^aD_a\phi],\\
        V_x&=\frac{8\pi}{\kappa\gamma}(\bar{A}'_{\rho}E^{\rho}-\bar{A}_xE^x{}')+4\pi P\partial_x\phi+4\pi\mathcal{A}\partial_x\pi^x,\\
        J&=-4\pi\partial_x\pi^x,
    \end{split}
\end{equation}
where $\tilde{K}:=K^i_aE^a_i=(E^x\bar{A}_x+E^{\rho}\bar{A}_{\rho})\sin\theta/\gamma$. The Poisson brackets between constraints can be calculated as:
\begin{equation}
    \begin{split}
        &\{J[A_0],J[A_0']\}=0,\quad \{J[A_0],H_x[N^x]\}=-J[\mathcal{L}_{N^x}A_0],\quad \{J[A_e],H_N[N]\}=0,\\
        &\{V_x[M^x],V_x[N^x]\}=-V_x[\mathcal{L}_{N^x}M^x],\quad \{H[N],V_x[N^x]\}=-H[\mathcal{L}_{N^x}N],\\
        &\{H[N],H^N[M]\}=V_x[(NM'-MN')E^2_x/|q|].
    \end{split}
\end{equation}
It is thus verified that after a number of canonical tranformations and simplifications of the original phase space, the theory remains an first-class constrained system.
\section{Deparametrization of The Symmetry-Reduced Theory}
In the previous section, we have obatined the 4-dimensional connection-dynamics of the spherically symmetric 5D Kaluza-Klein theory, and the constraints of the system are given in (\ref{spherical1}). In this section, we will solve the Gaussian constraint of the EM field and deparametrize the diffeomorphism constraint and the Hamiltionian constraint subsequently. First, by the deparametrization of the diffeomorphism constraint, the Hamiltionian constraint can be rewritten in terms of diffeomorphism invariant observables. We shall then deparametrize the Hamiltonian constraint and define relational evolution by explicitly calculating the physical Hamiltonian.
\subsection{Solving the Gaussian Constraint of the EM Field}
Since we have already imposed the Gaussian constraint for the 4-dimensional gravity during the spherically symmetric reduction, the remaining Hamiltonian constrained system, as is given in (\ref{spherical1}), contains three secondary first-class constraints. Since the Gaussian constraint of the EM field is not in deparametrizable form, we will first solve it before deparametrizing the remaining constraints.

It is obvious that the Gaussian constraint $J=-\partial_x\pi^x$ requires $\pi^x=constant$, and the only non-trivial gauge transformation corresponding to this constraint is:
\begin{equation}\label{gaugeA}
    \begin{split}
        e^{\{G_{EM}[N_0],\cdot\}}\mathcal{A}(x)=\mathcal{A}(x)-\partial_xN_0(x).
    \end{split}
\end{equation}
Let $\mathcal{A}(x)$ be an integrable function of $x$. Eq.(\ref{gaugeA}) implies that one can always choose the parameter function $N_0(x)$ of the gauge transformation such that $\mathcal{A}(x)=0$ by the gauge transformation. Note that, although the EM connection $\mathcal{A}(x)$ can be gauge fixed to be zero in the spherically symmetric case, the electric field $\mathcal{E}$ is nonzero. Using the definition for momentum $\pi^a$, the electric field can be expressed as:
\begin{equation}
    \mathcal{E}:=n^{\mu}F_{\mu x}=-\frac{\kappa E^{\rho}\pi^x}{\phi^3|E^x|^{\frac{3}{2}}}.
\end{equation}
After solving the Gaussian constraint for EM field, the only two remaining constraints are the following Hamiltonian constraint and diffeomorphism constraint in the radial direction:
\begin{equation}\label{spherical}
    \begin{split}
        H&=8\pi\{\frac{\kappa}{12E^{\rho}\sqrt{|E^x|}}P^2\phi+\frac{1}{3E^{\rho}\sqrt{|E^x|}}P\tilde{K}-\frac{\phi E^{\rho}}{\kappa\sqrt{|E^x|}}-\frac{\bar{A}^2_{\rho}E^{\rho}}{4\kappa\gamma^2\phi\sqrt{|E^x|}}-sgn(E^x)\frac{\sqrt{|E^x|}\bar{A}_{\rho}\bar{A}_x}{\kappa\gamma^2\phi}\\
        &\qquad +sgn(E^x)\frac{\phi\sqrt{|E^x|}}{\kappa}\partial_x(\frac{E^x{}'}{E^{\rho}})+\frac{\phi(E^x{}')^2}{4\kappa\sqrt{|E^x|}E^{\rho}}+\frac{\tilde{K}^2}{3\kappa\phi E^{\rho}\sqrt{|E^x|}}-\frac{\kappa\pi^x\pi^x E^{\rho}}{\phi^3\sqrt{|E^x|}|E^x|}+\frac{E^{\rho}\sqrt{|E^x|}}{\kappa}D^aD_a\phi\}=0,\\
        V_x&=\frac{8\pi}{\kappa\gamma}(\bar{A}'_{\rho}E^{\rho}-\bar{A}_xE^x{}')+4\pi P\partial_x\phi=0.\\
    \end{split}
\end{equation}

\subsection{Deparametrization of the Diffeomorphism Constraint}

On the constraint surface, by assuming that $E^x{}'$ is nonzero the diffeomorphism constraint can be written as:
\begin{equation}
    \bar{V}_x{}=\frac{8\pi}{\kappa\gamma}(\bar{A}_x-\frac{\bar{A}'_{\rho}E^{\rho}+\frac{\kappa\gamma}{2}P\partial_x\phi}{E^x{}'})=0,
\end{equation}
where $E^x$ is chosen as the reference field for the deparametrization. Then for a phase space function $f$, the corresponding diffeomorphism invariant observables can be constructed as:
\begin{equation}
    \begin{split}
        O_{f}(\sigma)=e^{\{\bar{V}_x{}[\beta^x],\cdot\}}f|_{E^x=l_p \sigma},
    \end{split}
\end{equation}
where $l_p$ is the Planck length. Here we have taken into account of the fact that the dimension of $E^x$ is $L^2$ \cite{chiou2012loop,campiglia2007loop}, while we want to choose the coordinate $\sigma$ to be of dimension $L$.

Now in the original phase space, before deparametrizing the diffeomorphism constraint, we consider the following phase space functions:
\begin{equation}\label{change1}
    \begin{split}
        \tilde{E}^{\rho}:=\frac{E^{\rho}}{E^x{}'},\quad \tilde{P}:=\frac{P}{E^x{}'}.
    \end{split}
\end{equation} 
It is easy to see that $\tilde{E^{\rho}}(x)$ and $\tilde{P}(x)$ are spatial scalar fields under the coordinate transformation along $x$ direction \cite{chiou2012loop}. Since $\bar{A}(x)$ and $\phi(x)$ are also spatial scalar fields, the diffeomorphism invariant observables of $\bar{A}_{\rho}(x)$, $\tilde{E}^{\rho}(x)$, $\phi(x)$ and $\tilde{P}(x)$ can be directly obtained by performing the coordinate transformation $x\mapsto\sigma$, i.e. 
\begin{equation}\label{OE}
    \begin{split}
        &O_{\bar{A}_{\rho}}(\sigma)=\bar{A}_{\rho}(\sigma),\quad O_{\tilde{E}^{\rho}}(\sigma)=\tilde{E}^{\rho}(\sigma),\quad O_{\phi}(\sigma)=\phi(\sigma),\quad O_{\tilde{P}}(\sigma)=\tilde{P}(\sigma).
    \end{split}
\end{equation}
Then the following Poisson brackets between diffeomorphism-invariant observables can be obtained by direct calculations:
\begin{equation}
    \begin{split}
        &\{O_{\bar{A}_{\rho}}(\sigma),O_{\tilde{E}_{\rho}}(\sigma')\}=\frac{\kappa\gamma}{8\pi l_p}\delta(\sigma,\sigma')\\
        &\{O_{\phi}(\sigma),O_{\tilde{P}_{\phi}}(\sigma')\}=\frac{1}{4\pi l_p}\delta(\sigma,\sigma').
    \end{split}
\end{equation}
Therefore, in the reduced phase space where the diffeomorphism constraint has been deparametrized, $\tilde{E}^{\rho}(\sigma)$ and $\tilde{P}(\sigma)$ are the conjugate momenta of $\bar{A}_{\rho}(\sigma)$ and $\phi(\sigma)$ respectively.

Recall that in the spherically reduced theory the components of 3-metric can be expressed by the densitized triad as:
\begin{equation}
    q_{xx}=\frac{(E^{\rho})^2}{|E^x|},\qquad q_{\theta\theta}=|E^x|,\qquad q_{\phi\phi}=|E^x|\sin^2\theta.
\end{equation}
Since $|E^x|$ is associated with the square of curvature radius at point $x$, for the Schwarzchild spacetime it changes monotonically along the radial directon. This motivates us to choose $E^x$ as the reference field for the spatial direction.

\subsection{Deparametrization of the Hamiltionian Constraint}

Consider the intermediate Hamiltonian $\bar{H}:=\frac{H}{E^x{}'}$, which transforms as a spatial scalar field in the spherically symmetric model. Then its diffeomorphism invariant observable $O_{\bar{H}}(\sigma)$, just like $O_{\bar{A}_{\rho}}(\sigma)$ and $O_{\tilde{E}^{\rho}}(\sigma)$ in (\ref{OE}), can also be viewed as a coordinate transformation of $\bar{H}(x)$ from $x$ to $\sigma$. Given a suitable function $\tilde{N}(\sigma)$, the smeared version of the diffeomorphism invariant intermediate Hamiltonian can be defined as:
\begin{equation}
    \begin{split}
        \bar{H}[\tilde{N}]&:=\frac{8\pi}{\kappa}\int d\sigma \tilde{N}(\sigma)\bar{H}(\sigma).
    \end{split}
\end{equation}
Note that, because of Eq.(\ref{ftof}), from now on we use $\bar{H}(\sigma)$ to represent $O_{\bar{H}}(\sigma)$ and the same notation applies for all variables in the diffeomorphism invariant phase space as well.

Recall that the scalar $\phi$ in the original Kaluza-Klein theory represents the circumference of the fifth dimension. Therefore, in order for $\phi$ to be chosen as the reference field for time evolution, we shall assume that $\phi$ changes monotoniacally in time. Thus we consider the cases where the fifth dimension is either expanding or shrinking.

Using Eqs.(\ref{spherical}) and (\ref{change1}), the intermidiate Hamitonian can be written as:
\begin{equation}
    \begin{split}
        \bar{H}(\sigma)&=8\pi\{[\frac{\kappa\phi}{12\sqrt{|l_p\sigma|}\tilde{E}^{\rho}}+\frac{\kappa |\sigma|^{\frac{3}{2}}(\partial_{\sigma}\phi)^2}{12\sqrt{l_p}\phi\tilde{E}^{\rho}}+\frac{\kappa sgn(\sigma)\sqrt{|\sigma|}\partial_{\sigma}\phi}{6\sqrt{l_p}\tilde{E}^{\rho}}]\tilde{P}^2\\
        &+[\frac{|\sigma|^{\frac{3}{2}}\partial_{\sigma}\bar{A}_{\rho}\partial_{\sigma}\phi}{3\sqrt{l_p}\gamma\phi}+\frac{sgn(\sigma)\sqrt{|\sigma|}\partial_{\sigma}\bar{A}_{\rho}}{3\sqrt{l_p}\gamma}-\frac{sgn(\sigma)\sqrt{|\sigma|}\bar{A}_{\rho}\partial_{\sigma}\phi}{6\sqrt{l_p}\phi\gamma}+\frac{\bar{A}_{\rho}}{3\gamma\sqrt{l_p|\sigma|}}]\tilde{P}\\
        &+\frac{\phi sgn(\sigma)\sqrt{|\sigma|}}{\sqrt{l_p}\kappa}\partial_{\sigma}(\frac{1}{\tilde{E}^{\rho}})+\frac{\phi}{4\kappa\sqrt{l_p|\sigma|}\tilde{E}^{\rho}}-sgn(\sigma)\frac{\sqrt{|\sigma|}\bar{A}_{\rho}\tilde{E}^{\rho}\partial_{\sigma}\bar{A}^{\rho}}{\sqrt{l_p}\phi\kappa\gamma^2}-\frac{\kappa\pi^x\pi^x\tilde{E}^{\rho}}{\phi^3(l_p\sigma)^2}\\
        &+\frac{|\sigma|^{\frac{3}{2}}(\partial_{\sigma}\bar{A}_{\rho})^2\tilde{E}^{\rho}}{3\sqrt{l_p}\kappa\gamma^2\phi}+(\frac{3sgn(\sigma)\sqrt{|\sigma|}}{2\sqrt{l_p}\tilde{E}^{\rho}}-\frac{|\sigma|{}^{\frac{3}{2}}}{\sqrt{l_p}(\tilde{E}^{\rho})^2}\partial_{\sigma}\tilde{E}^{\rho})\partial_{\sigma}\phi\\
        &+\frac{|\sigma|{}^{\frac{3}{2}}}{\sqrt{l_p}\tilde{E}^{\rho}}\partial_{\sigma}\partial_{\sigma}\phi-\frac{\phi\tilde{E}^{\rho}}{\kappa\sqrt{l_p|\sigma|}}-\frac{\bar{A}^2_{\rho}\tilde{E}^{\rho}}{4\kappa\phi\gamma^2\sqrt{l_p|\sigma|}}+\frac{\tilde{E}^{\rho}\bar{A}_{\rho}^2+2\bar{A}_{\rho}\sigma\partial_{\sigma}\bar{A}_{\rho}\tilde{E}^{\rho}}{3\kappa\gamma^2\phi\sqrt{l_p|\sigma|}}\}.\\
    \end{split}
\end{equation}
Thus the constraint $\bar{H}=0$ can be seen as a quadratic equation with respect to $\tilde{P}$, which takes the form $a\tilde{P}^2+b\tilde{P}+c=0$, where the three coefficients can be written as:
\begin{equation}
    \begin{split}
        a&:=2\pi\frac{\kappa(\phi+\sigma\partial_{\sigma}\phi)^2}{3\sqrt{|l_p\sigma|}\tilde{E}^{\rho}\phi},\\
        b&:=\frac{4\pi}{3\gamma\phi}(2\frac{|\sigma|^{\frac{3}{2}}}{\sqrt{l_p}}\partial_{\sigma}\bar{A}_{\rho}\partial_{\sigma}\phi+2sgn(\sigma)\phi\sqrt{\frac{|\sigma|}{l_p}}\partial_{\sigma}\bar{A}_{\rho}-sgn(\sigma)\sqrt{\frac{|\sigma|}{l_p}}\bar{A}_{\rho}\partial_{\sigma}\phi+\frac{2\phi\bar{A}_{\rho}}{\sqrt{l_p|\sigma|}}),\\
        c&:=\bar{H}-a(\pi_{\phi})^2-b\pi_{\phi}.\\
    \end{split}
\end{equation}
In the special case when $a=0$, one has $\partial_{\sigma}\phi=-\frac{\phi}{\sigma}$ and $b=\frac{4\pi \bar{A}_{\rho}}{\gamma\sqrt{l_p|\sigma|}}$. Then we have the following two special cases. First, if $b=0$, one gets $\bar{A}_{\rho}=0$ and $c=0$. Since the only remaining phase space variable present in $c$ is $\tilde{E}^{\rho}$, this case corresponds to a special solution of the theory after imposing the Hamiltonian constriant. Second, if $b\neq 0$, the model is deparametrizable. The Hamiltonian constraint becomes:
\begin{equation}
    \bar{H}(\sigma)=b\tilde{P}+c=0.
\end{equation}
Hence one obtains
\begin{equation}
    \tilde{P}+\frac{\gamma\sqrt{l_p|\sigma|}c}{4\pi \bar{A}_{\rho}}=0,
\end{equation}
which is suitable for deparametrization by using $\phi$ as the reference field for time. The construction of gauge-invariant observables can be done following the framework introduced in section II. We will show the deparametrization process explicitly below for the more general case when $a\neq 0$. In this case, on the constraint surface the condition $b^2-4ac\ge 0$ is automatically satisfied. Therefore, we can reformulate the Hamiltonian constraint in the following form:
\begin{equation}
    \tilde{H}=\tilde{P}+(H_1\pm H_2),
\end{equation}
where:
\begin{equation}
    \begin{split}
        H_1(\sigma)&=\frac{2\tilde{E}^{\rho}\sigma^{2}\partial_{\sigma}\bar{A}_{\rho}\partial_{\sigma}\phi+2\phi\tilde{E}^{\rho}\sigma\partial_{\sigma}\bar{A}_{\rho}-\tilde{E}^{\rho}\sigma\bar{A}_{\rho}\partial_{\sigma}\phi+2\phi\tilde{E}_{\rho}\bar{A}_{\rho}}{\kappa\gamma(\phi+2\sigma\partial_{\sigma}\phi)^2},\\
        H_2(\sigma)&=-\frac{\sqrt{l_p|\sigma|}\tilde{E}^{\rho}}{\kappa\gamma(\phi+2\sigma\partial_{\sigma}\phi)^2}\sqrt{H_s},\\
    \end{split}
\end{equation}
with:
\begin{equation}
    \begin{split}
        H_s&=\frac{4}{l_p}|\sigma|^3(\partial_{\sigma}\bar{A}_{\rho}\partial_{\sigma}\phi)^2+\frac{4}{l_p}|\sigma|(\partial_{\sigma}\bar{A}_{\rho})^2\phi^2+\frac{|\sigma|}{l_p}\bar{A}_{\rho}^2(\partial_{\sigma}\phi)^2+\frac{4\phi^2\bar{A}_{\rho}{}^2}{l_p|\sigma|}\\
        &+\frac{8}{l_p}\sigma|\sigma|(\partial_{\sigma}\bar{A}_{\rho})^2\phi\partial_{\sigma}\phi-\frac{4}{l_p}|\sigma|\bar{A}_{\rho}\partial_{\sigma}\bar{A}_{\rho}\partial_{\sigma}\phi-\frac{4}{l_p}\sigma|\sigma|\bar{A}_{\rho}\phi\partial_{\sigma}\bar{A}_{\rho}(\partial_{\sigma}\phi)^2\\
        &+\frac{8}{l_p}\phi|\sigma|\bar{A}_{\rho}\partial_{\sigma}\bar{A}_{\rho}\partial_{\sigma}\phi+\frac{8}{l_p}sgn(\sigma)\phi^2\bar{A}_{\rho}\partial_{\sigma}\bar{A}_{\rho}-\frac{4}{l_p}\phi sgn(\sigma)\bar{A}_{\rho}{}^2\partial_{\sigma}\phi\\
        &-12\kappa\phi\gamma^2(\phi+2\sigma\partial_{\sigma}\phi)^2\\
        &\times(\frac{l_p\phi sgn(\sigma)}{\kappa\tilde{E}^{\rho}}\partial_{\sigma}(\frac{1}{\tilde{E}^{\rho}})+\frac{\phi}{4l_p\kappa|\sigma|(\tilde{E}^{\rho})^2}-sgn(\sigma)\frac{\bar{A}_{\rho}\partial_{\sigma}\bar{A}^{\rho}}{l_p\phi\kappa\gamma^2}-\frac{\kappa\pi^x\pi^x}{l_p\phi^3\sigma^2\sqrt{|\sigma|}}\\
        &+\frac{|\sigma|(\partial_{\sigma}\bar{A}_{\rho})^2}{3l_p\kappa\gamma^2\phi}+(\frac{3sgn(\sigma)}{2l_p(\tilde{E}^{\rho})^2}-\frac{|\sigma|}{l_p(\tilde{E}^{\rho})^3}\partial_{\sigma}\tilde{E}^{\rho})\partial_{\sigma}\phi\\
        &+\frac{|\sigma|}{l_p(\tilde{E}^{\rho})^2}\partial_{\sigma}\partial_{\sigma}\phi-\frac{\phi}{l_p\kappa|\sigma|}-\frac{\bar{A}^2_{\rho}}{4l_p\kappa\phi\gamma^2|\sigma|}+\frac{\bar{A}_{\rho}^2+2\bar{A}_{\rho}\sigma\partial_{\sigma}\bar{A}_{\rho}}{3l_p\kappa\gamma^2\phi|\sigma|}).\\
    \end{split}
\end{equation}
Given an arbitrary phase space function $f$, its gauge-invariant Dirac observable invariant with respect to the Hamiltonain constraint can be constructed as:
\begin{equation}
    \begin{split}
        f(\tau,\sigma):=O_{O_{f}(\sigma)}(\tau)=e^{\{\tilde{H}{}[\beta],\cdot\}}(O_{f}(\sigma))|_{\phi=\tau},
    \end{split}
\end{equation}
where $\tau$ is chosen to be independent of $\sigma$, i.e., $\partial_{\sigma}\tau=0$, so that we have completely separated spatial and time coordinates. Physically, this setting implies that the physical clocks on different points on each spatial hypersurface are synchronized to have a single value $\tau$. On the final reduced phase space, only one pair of physical DoFs remains and they satisfy:
\begin{equation}
    \begin{split}
        &\{O_{\bar{A}_{\rho}}(\tau,\sigma),O_{\tilde{E}_{\rho}}(\tau,\sigma')\}=\frac{\kappa\gamma}{8\pi l_p}\delta(\sigma,\sigma').\\
    \end{split}
\end{equation}

After deparametrizing the Hamiltonian constraint, we are left with the following smeared physical Hamiltonian which determines the physical time evolution of the system:
\begin{equation}
    O_{H[1]}(\tau,\sigma)=H_1(\tau,\sigma)\pm H_2(\tau,\sigma),
\end{equation}
where:
\begin{equation}
    \begin{split}
    H_1(\tau,\sigma)&=\frac{2\tilde{E}^{\rho}\sigma\partial_{\sigma}\bar{A}_{\rho}+2\tilde{E}_{\rho}\bar{A}_{\rho}}{\kappa\gamma\tau},\\
    H_2(\tau,\sigma)&=-\frac{\sqrt{|\sigma|}\tilde{E}^{\rho}}{l_p\kappa\gamma\tau^2}\sqrt{H_s(\tau,\sigma)},\\
    H_s(\tau,\sigma)&=-12\kappa\gamma^2\tau^3\times(\frac{\tau sgn(\sigma)}{l_p\kappa\tilde{E}^{\rho}}\partial_{\sigma}(\frac{1}{\tilde{E}^{\rho}})+\frac{\tau}{4l_p\kappa|\sigma|(\tilde{E}^{\rho})^2}-sgn(\sigma)\frac{\bar{A}_{\rho}\partial_{\sigma}\bar{A}^{\rho}}{l_p\tau\kappa\gamma^2}-\frac{\kappa\pi^x\pi^x}{l_p^2\tau^3\sigma^2\sqrt{l_p|\sigma|}}\\
    &-\frac{\tau}{l_p\kappa|\sigma|}-\frac{\bar{A}^2_{\rho}}{4l_p\kappa\tau\gamma^2|\sigma|}).
    \end{split}
\end{equation}
Finally, the evolution equation of basic variables can be obtained by using Eq.(\ref{evolution}) as:
\begin{equation}
    \begin{split}
        \frac{\partial\tilde{E}^{\rho}(\tau,\sigma)}{\partial \tau}&=\{O_{h[1]}(\tau),\tilde{E}^{\rho}(\tau,\sigma)\}\\
        &=-\frac{\kappa\gamma}{8\pi}[\frac{2\sigma\partial_{\sigma}\tilde{E}^{\rho}}{\kappa\gamma\tau}\pm(-\frac{6}{l_p\kappa\gamma}\partial_{\sigma}(\frac{sgn(\sigma)\sqrt{l_p|\sigma|}\tilde{E}^{\rho}}{\sqrt{H_s}})+\frac{3\tilde{E}^{\rho}}{l_p\kappa\gamma\sqrt{|\sigma|}\sqrt{H_s}})\bar{A}_{\rho}],\\
        \frac{\partial\bar{A}_{\rho}(\tau,\sigma)}{\partial \tau}&=\{O_{h[1]}(\tau),\bar{A}_{\rho}(\tau,\sigma)\}\\
        &=-\frac{\kappa\gamma}{8\pi}[\frac{2\sigma\partial_{\sigma}\bar{A}_{\rho}+2\bar{A}_{\rho}}{\kappa\gamma\tau}\mp(\frac{\sqrt{l_p|\sigma|}\sqrt{H_s}}{\kappa\gamma\tau^2}+\frac{3\gamma\tau^2}{l_p\kappa}\partial_{\sigma}(\frac{sgn(\sigma)\sqrt{l_p|\sigma|}\tilde{E}^{\rho}}{\sqrt{H_s}})\frac{2}{(\tilde{E}^{\rho})^3}+\frac{3\gamma\tau^2}{\kappa\sqrt{l_p|\sigma|}\sqrt{H_s}(\tilde{E}^{\rho})^2})].
    \end{split}
\end{equation}

\subsection{The $\pm$ Sign Problem}

One ambiguity still remains for our deparametrization scheme: A $\pm$ sign appears when we rewrite the intermediate Hamiltonian into its deparametrized form. Now we clarify this ambiguity by considering the homogeneous and isotropic cosmology in the Kaluza-Klein theory. Recall that the induced line element of the spherically symmetric metric on an 3-dimensional spacelike hypersurface can be written in terms of the spherical coordinates $x^a=(x,\theta,\varphi)$ by two functions $\Lambda(t,x)$ and $R(t,x)$ as:
\begin{equation}\label{metric1}
    ds^2=q_{ab}dx^adx^b=\Lambda^2(t,x)dx^2+R^2(t,x)d\Omega^2,
\end{equation}
where $d\Omega^2=d\theta^2+\sin^2\theta d\phi^2$, and the metric components are related to those of the densitized triad by:
\begin{equation}
    |E^x|=R^2,\qquad E^{\rho}=R\Lambda.
\end{equation}
Since the line element of the Friedman-Lemaître-Robertson-Walker (FLRW) metric
\begin{equation}
    \begin{split}
        ds=-dt^2+a^2(t)[dx^2+x^2(d\theta^2+\sin\theta d\varphi^2)]
    \end{split}
\end{equation}
is a special case of the spherically symmetric one, we can obtain the expressions of the basic variables in the FLRW model as:
\begin{equation}\label{lqc1}
    \begin{split}
        &N=1,\quad N^a=0,\quad \Lambda=a,\quad R=xa,\\
        &E^{\rho}=xa^2,\quad E^x=a^2x^2,\quad \bar{A}_{\rho}=2\gamma x\dot{a},\quad \bar{A}_x=\gamma\dot{a},\\
        &\tilde{K}=3 x^2a^2\dot{a},\quad P=\frac{3x^2a^3}{\kappa}\frac{\dot{\phi}}{\phi}-\frac{6 x^2a^2\dot{a}}{\kappa\phi}.
    \end{split}
\end{equation}
Note that the expressions in (\ref{lqc1}) coincide with those in the equivalent model of Brans-Dicke theory \cite{zhang2013loop}. While the diffeomorphism constraint is satisfied automatically in the homogeneous model, the Hamiltonian constraint remains to be deparametrized. It can be recast into the following form:
\begin{equation}\label{vacuum}
    \begin{split}
        \tilde{H}:=&P+(H_1\pm H_2)=0,\\
    \end{split}
\end{equation}
where:
\begin{equation}\label{H123}
    \begin{split}
        H_1&=\frac{2\tilde{K}}{\kappa\phi},\quad H_2=-\sqrt{H_s},\\
        H_s&=\frac{12E^{\rho}{}^2}{\kappa^2}+\frac{3\bar{A}_{\rho}^2E^{\rho}{}^2}{\gamma^2\kappa^2\phi^2}+\frac{12\bar{A}_{\rho}\bar{A}_xE^{\rho}{}E^x}{\gamma^2\kappa^2\phi^2}-\frac{12E^{\rho}{}E^x}{\kappa^2}\partial_x(\frac{E^x{}'}{E^{\rho}})-\frac{3(E^x{}')^2}{\kappa^2}.
    \end{split}
\end{equation}
By Eqs. (\ref{lqc1}), the following equation can be derived from (\ref{vacuum}):
\begin{equation}\label{friedman}
    \frac{\dot{\phi}}{\phi}=\pm\sqrt{(\frac{2\dot{a}}{a}+\frac{\dot{\phi}}{\phi})^2}.
\end{equation}
Let us first consider the case of taking the "plus" sign in Eq.(\ref{friedman}). Then if $\frac{\dot{\phi}}{\phi}<-\frac{2\dot{a}}{a}$ the equation becomes $\frac{\dot{a}}{a}+\frac{\dot{\phi}}{\phi}=0$, which is exactly the Friedman equation for the homogeneous and isotropic spacetime in 5-dimensional Kaluza-Klein theory. If $\frac{\dot{\phi}}{\phi}\geq-\frac{2\dot{a}}{a}$, only the trivial solution $\dot{a}=0$ can be obtained. Another case is to take "minus" sign in Eq.(\ref{friedman}). Then if $\frac{\dot{\phi}}{\phi}<-\frac{2\dot{a}}{a}$, one gets the trivial solution $\dot{a}=0$. If $\frac{\dot{\phi}}{\phi}\geq-\frac{2\dot{a}}{a}$, we have again $\frac{\dot{a}}{a}+\frac{\dot{\phi}}{\phi}=0$. Hence the two non-trivial sectors shares the same equation of motion. Moreover, it is straightforward to check that, by performing the deparametrization of the Hamiltonian constraint, the physical Hamitonians obtained in both cases also take the same form.
Therefore, both the "plus" sign theory and the "minus" sign theory contain non-trivial solutions belonging to different sectors in the phase space. Hence both cases should be taken into account when the deparametrized theory is considered.

\section{Concluding Remark}

In the previous sections, the Hamiltonian analysis of the 5-dimensional Kaluza-Klein theory reduced on 4-dimensional spacetime has been performed. Its connection dynamics has been obtained by canonical transformations. Then the spherically symmetric model of 
the theory is studied. We have successfully defined relational evolution for this higher dimensional gravity theory in its spherically symmetric model by using only the gravitational degrees of freedom. Conceptually speaking, in such a system physical time evolution and spatial coordinates are depicted by the geometrical degrees of freedom of the higher dimensional spacetime, rather than being depicted by matter fields introduced additionally to the gravity theory. The fields we used to deparametrize the Hamiltionian constraint and diffeomorphism constraint are the scalar field $\phi$ and the radial component $E^x$ of the densitized triad respectively, where $\phi$ is obtained via killing reduction of the scale of the fifth dimension in 5-dimensional Kaluza-Klein theory.

It should be noted that the full deparametrization of the general 4-dimensional cosntrained system of the 5-dimensional Kaluza-Klein theory has not been achieved in this work, while the deparametrization of the spherically symmetric model has been done. In the full theory it is very difficult to recast the cosntrained system into a form suitable for the deparametrization by its own degrees of freedom. Nevertheless, the deparametrization of the model indicates the scheme to that of the full theory. It is worth studying further the quantization of this deparametrized model, which we leave for our next work. It is also expected that our classical deparametrization of the theory by its geometrical degrees of freedom could provide some hints on understanding the issues of quantum time and quantum space in quantum gravity.

\section*{Acknowledgements}
This work is supported by the National Natural Science Foundation of China (Grant Nos. 11875006, and 11961131013). We thank Hongguang Liu, Gaoping Long, Shupeng Song, Faqiang Yuan and Cong Zhang for helpful discussions.





\end{document}